\documentclass[prd,showpacs,nofootinbib,superscriptaddress,twocolumn,preprintnumbers]{revtex4}

\usepackage[mathscr]{eucal}
\usepackage[latin1]{inputenc}
\usepackage{amsmath}
\usepackage{amsfonts}
\usepackage{amssymb}
\usepackage{graphicx}

\newcommand{\R}{\mathbb{R}}

\def\pmk{k}
\def\Hm{{H_{\mathrm{m}}}} 

\def\H{{\cal H}}

\def\g{\gamma}

\def\lp{{\ell}_{\rm Pl}}

\def\R{\mathbb{R}}

\newcommand{\rcr}{\rho_{\mathrm{crit}}}

\newcommand{\heff}{{\cal H}_{\mathrm{eff}}}

\newcommand{\hcalm}{{\cal H}_{\mathrm{M}}}

\newcommand{\f}{\frac}

\newcommand{\fref}[1]{Fig.\,\ref{#1}}

\def\sfnsq{\sin^2\bar \mu}
\def\sfnsqc{\sin^2\,\big(\mb(c - k)\big)}

\def\sk{\sin\big(\mb(c\! - \! k)\!\big)}
\def\ck{\cos\big(\mb(c\! - \! k)\!\big)}
\def\smk{\sin^2\big(\mb(c \! - \!  k)\!\big)}

\usepackage{enumerate}

\usepackage{colordvi}
\newcounter{mnotecount}[section]

\newcommand{\comment}[1]{}
\newcommand{\fs}[2]{{\textstyle\frac{#1}{#2}}} 

\def\f{\frac}

\def\d{\textrm{d}}

\def\mb{\bar \mu}

\usepackage{enumerate}

\newcommand{\be}{\nopagebreak[3]\begin{equation}}
\newcommand{\ee}{\end{equation}}
\newcommand{\ba}{\nopagebreak[3]\begin{eqnarray}}
\newcommand{\ea}{\end{eqnarray}}
\newcommand{\nn}{\nonumber \\}

\def\pmk{k}
\def\Hm{{H_{\mathrm{m}}}} 

\def\H{{\cal H}}

\def\g{\gamma}

\def\lp{{\ell}_{\rm Pl}}

\def\R{\mathbb{R}}

\def\f{\frac}

\def\d{\textrm{d}}

\def\mb{\bar \mu}

\usepackage{enumerate}
\begin{document}

\title{Exotic singularities and spatially curved Loop Quantum Cosmology}

\author{Parampreet Singh}
\email{psingh@phys.lsu.edu}
\affiliation{Department of Physics and Astronomy, Louisiana State University,
Baton Rouge, Louisiana 70803, USA}
\affiliation{Perimeter Institute for
Theoretical Physics, 31 Caroline Street North, Waterloo, Ontario
N2L 2Y5, Canada}

     \author{Francesca Vidotto}
 \email{vidotto@cpt.univ-mrs.fr}
     \affiliation{Centre de Physique Th\'eorique de Luminy
     , Case 907, F-13288 Marseille, EU}
     \affiliation{Dipartimento di Fisica Nucleare e Teorica,
        Universit\`a degli Studi di Pavia \\ and Istituto Nazionale
        di Fisica Nucleare, Sezione di Pavia, via A. Bassi 6,
        27100 Pavia, EU}
        

\begin{abstract}        
We investigate the occurrence of various exotic spacelike singularities in the past and the future evolution of $k = \pm 1$  Friedmann-Robertson-Walker model and loop quantum cosmology using a sufficiently general phenomenological model for the equation of state. 
We highlight the non-trivial role played by  the intrinsic curvature for these singularities and the new physics which emerges at the Planck scale. We show that quantum gravity effects generically resolve all strong 
curvature singularities including big rip and big freeze singularities. The weak singularities, which include sudden and big brake singularities are ignored by quantum gravity when spatial curvature is negative, as was previously found for the spatially flat model. Interestingly, for the spatially closed model there exist cases where weak singularities may be resolved when they occur in the past evolution. The spatially closed model exhibits another novel feature. 
For a particular class of equation of state, 
this model also exhibits  an additional physical branch in loop quantum cosmology, a baby universe separated from the parent branch. Our analysis generalizes previous results obtained on  the resolution of  
strong curvature singularities in flat models to isotropic spacetimes with non-zero spatial curvature.

\end{abstract}

\maketitle


\section{Introduction}

A fundamental limitation of the cosmological models based on general relativity (GR) is the occurrence of singularities. Perhaps one of the simplest examples is the case of an expanding homogeneous and isotropic universe filled with a matter satisfying strong energy condition such as dust or radiation. Independent of the intrinsic geometry of the universe, be it  closed, flat or open, the past evolution of such a universe from arbitrary initial conditions leads to an initial singularity: the big bang, where the classical 
dynamical equations break down and the physics stops. Another example is the case of inflationary universe in which even though the
evolution is almost deSitter, the classical spacetime is past incomplete \cite{bgv}.

In recent years, various new singularities have been found in classical cosmology \cite{bigrip,sudden,sudden1,future}. Unlike the 
big bang and big crunch singularities, these singularities do not occur when scale factor vanishes. 
These occur either at finite values of the scale factor or when it diverges. Recall that for the matter satisfying 
weak energy condition, the latter is not possible as the spacetime curvature goes to zero when
scale factor goes to infinity. However, if matter violates weak energy condition, for example in the case of phantom fields, then spacetime 
curvature will diverge as scale factor becomes infinite. For homogeneous and isotropic models with matter equation of state in the form of a perfect fluid these exotic singularities come in four types. Big rip (type I) where 
the energy density and pressure diverge along with a divergence in the scale factor, sudden singularity (type II) occurring at a finite value of the scale factor and energy density with a divergence in pressure, big freeze (type III) where energy density and pressure diverge at a finite value of the scale factor and big brake (type IV) singularity where scale factor, energy density and pressure are finite  but there is a divergence in the time derivative of the pressure or rate of change of energy density.

Lack of  successful resolution of these singularites renders classical cosmological models incomplete. It is generally believed that existence of these singularities is a result of assuming 
the validity of GR even in the regime of large spacetime curvature where the effects due to quantum gravity may become important and lead to significant departures 
from the classical theory. It is thus hoped that incorporation of quantum gravitational effects may result  in a possible resolution of these singularities.

Loop quantum gravity (LQG) is one of the candidate theories of quantum gravity which attempts to address 
this issue. It is a non-perturbative 
and background independent quantization, with a key prediction that  the continuum differential geometry of the classical theory is replaced by a discrete quantum geometry in the quantum theory. 
Perhaps one of the best illustrations of the novel effects of quantum geometry is captured in loop quantum cosmology (LQC) which is a quantization of homogeneous spacetimes based on LQG \cite{ashtekar:lqc_review,bojowald:livingreview,singh:review1}. A key prediction of LQC is that 
the big bang singularity is replaced by a big bounce, which is a direct consequence of the underlying quantum geometry 
\cite{aps,aps1,aps2,apsv}. 
These results which were 
first obtained for homogeneous and isotropic models (for all values of spatial curvature) with a massless scalar field have been extended 
to inflationary potential 
\cite{aps3}, 
anisotropic spacetimes 
\cite{cv:bianchi1,aw:bianchi1}
and also certain inhomogeneous situations 
\cite{gowdy}.
Further, using an  exactly solvable model it has been shown that the expectation values of energy density have a universal upper bound for a dense subspace in the physical Hilbert space \cite{slqc}. There are strong constraints on the change in relative fluctuations of quantum observables across the bounce 
\cite{recall}. Recently, much stronger constraints on the change in dispersions have been obtained by Kaminski and Pawlowski \cite{polish_recall}. These results show that a universe like ours i.e. macroscopic at late times bounces from a 
a  similar universe at very early times (in the contracting branch) and the universe recalls almost of all its state through the bounce.

Interestingly, the loop quantum dynamics admits an effective description on a continuum spacetime 
which can be obtained using coherent state techniques \cite{vt,st}. An important feature of this analysis is that 
one can obtain an effective Hamiltonian from which one can obtain modified Friedmann and Raychaudhuri  
equations as the Hamilton's equations. The modified set of dynamical equations inherit quantum geometric effects via higher order non-perturbative corrections which vanish at small spacetime curvatures. 
It is important to note that various numerical simulations have shown that  effective equations capture the underlying quantum evolution very accurately for universes which become macroscopic at late times. 
These thus prove to be useful tools to understand the physics in LQC, such as whether the underlying theory has 
well defined ultra-violet and infra-red limits. It turns out that even though there exist various quantization ambiguities, there is a unique quantization leading to a consistent unambiguous physical description
\cite{cs1,cs2} (the improved dynamics \cite{aps2,apsv}: which is being considered here).

Using effective equations, we can ask various questions regarding the generality of singularity 
resolution in LQC. For example, one can ask whether spacetime curvature is always bounded in LQC? 
Here we should note that a universal bound on energy density (as in LQC), does not imply that the spacetime curvature is also bounded. This is easy to understand for the classical cosmology, where the Ricci scalar, which provides us a complete information about the spacetime curvature in the homogeneous and isotropic spacetime,
depends both on the energy density and pressure.  Though 
for most matter-energy configurations, the behavior of equation of state is such that an upper bound in 
energy density is sufficient to control the divergence in pressure and hence the spacetime curvature, it is not difficult to come up with counter examples with a more general equation of state \cite{portsmouth,generic}. Hence, an upper bound in energy density is not sufficient to prevent a divergence 
in the spacetime curvature.

A pertinent question is whether this divergence signals the end of spacetime in LQC. In order to answer this 
question, we recall that even in GR we encounter events where spacetime curvature blows up but there is no
associated physical singularity. This can happen if the tidal forces are not strong enough to cause a 
complete destruction of in-falling objects in to the singularity and geodesics can be extended beyond 
such events. It turns out that the events where spacetime curvature diverges in flat isotropic LQC are 
weak singularities and geodesics can be extended beyond them. In flat isotropic LQC, the divergence of spacetime curvature occurs only for sudden singularities which are caused by a divergence in pressure at a finite 
scale factor and energy density. It is straight forward to show that the expansion parameter in this case is bounded and the spacetime is geodesically complete in the flat isotropic LQC \cite{generic}.

In this paper we take the first step to generalize the above result by including intrinsic curvature in the spacetime. This is done by considering the effective dynamics of 
loop quantized spatially closed and open models in the Robertson-Walker geometry. In the classical Friedmann dynamics, intrinsic curvature term enters 
in form of $1/a^2$ term in the dynamical equations. Thus it is quite straightforward to understand  the expected modifications from the results in  the 
spatially flat model. In LQC,  the quantization of intrinsic curvature brings non-trivial modifications to the effective description and makes the resulting form of effective 
dynamical equations less straight forward to analyze. Though one expects that at small intrinsic curvatures one recovers the results of flat isotropic LQC, 
physics may be bring up surprises when intrinsic curvature is large. 
As we will show, this is indeed what happens in the case of $k=\pm 1$ models in LQC. 

Our analysis will be based on considering a sufficiently general phenomenological model for the equation of state which was proposed in Ref.\cite{not}. This model allows a study of all of the 
exotic singularities by the choice of different parameters and was earlier used for investigation of resolution of strong curvature singularities in the flat isotropic LQC \cite{generic}. As we will see, the effective dynamical equations of spatially curved models approximate those of the flat model in LQC at large volumes because the contribution from intrinsic curvature becomes negligible in this limit. Thus for future singularities, at large volumes, the resulting physics is similar 
for models with or without spatial curvature in LQC. However for certain values of parameters the spatially closed model in LQC permits two separate physical branches, a short lived baby universe at small volume and a parent universe which evolves to a macroscopic size. This branch is absent in the classical theory and has a pure quantum geometric origin.

To completely capture the new physics from inclusion of intrinsic curvature, it is important to 
study exotic singularities in the past evolution when they occur at small volumes. Our analysis 
of past and future exotic singularities shows that all strong curvature singularities are resolved 
in $k=\pm 1$ isotropic LQC. The scale factors at which big rip and big freeze singularities occur 
are excluded from the allowed range by loop quantum effects. As in the flat model, the spacetime curvature can diverge in spatially curved LQC, however when ever this happens one has a weak singularity which is known to be harmless. In almost all cases these singularities are ignored 
by LQC. The only exception to this occurs in spatially closed model where weak singularities occurring in the past evolution may be resolved. This occurs purely because of the non-trivial role 
of intrinsic curvature effects in LQC. 

\vspace{5pt}
We organize our paper as follows: In Sec.\ref{classical} we revisit the classical equations for spatially curved model in classical cosmology and introduce the phenomenological ansatz of the equation of state. To facilitate the reader to follow the derivation of effective equations in LQC, we derive the 
Friedmann and Raychaudhuri  equations in the Hamiltonian framework. In Sec.\ref{quantum}, we derive the effective equations for spatially curved LQC starting from the effective Hamiltonian 
\cite{apsv,kv,szulc}. 
Using these modified equations, we numerically obtain solutions and discuss the new physics in Sec.\ref{numerics}. Here we show that all exotic strong curvature singularities, irrespective of whether they occur in the past or the future, are resolved in spatially curved LQC. We summarize our results in Sec.\ref{summary}.

\section{Classical theory}\label{classical}

\subsection{Hamiltonian cosmology}
The fundamental equations of cosmology describe the evolution of the scale factor $a$ in proper time $t$.
These equations can be derived in a simple Hamiltonian framework.  To this purpose, consider the conjugate variables $(c,p)$ where $c$ is the Ashtekar-Barbero connection and $p$ is the triad (which without any loss of generality will be chosen with positive orientation). These are related to the metric variables as
\be\label{variables}c = \g\dot a +\pmk \hskip50pt   p = a^2,  \ee
with the relation between $c$ and $\dot a$ valid only in the classical theory and it modifies when quantum gravity effects are considered. Here $\dot a=\d a/\d t$, $k=0,\pm1$ is the normalized intrinsic curvature and $\g\in\R$ is the Barbero-Immirzi parameter%
\footnote{
The value of $\gamma$ can be fixed by computing the black hole entropy in LQG.
In our numerical analysis we set $\gamma\approx0.2375$ \cite{Meissner:2004ju}.
}%
. The conjugate variables satisfy the following Poisson bracket 

%
\be \{c,\, p\} = \frac{8 \pi G}{3}\g ~.  \ee
In these variables, the Hamiltonian for gravity in a homogeneous and isotropic spacetime becomes
\be\label{hclk} \H_g = -
\frac{3}{8 \pi G} \ 
\gamma^{-2 } \sqrt p  \, \left[ (c-\pmk)^2 +\pmk\g^2\right] ~.
\ee
We introduce a generic matter field with Hamiltonian
$
\H_m = p^{\f32} \rho
$, 
where  $\rho$ is the matter energy density. 
The total Hamiltonian of the system
\ 
$\H=\H_g+\H_m$
is constrained to vanish.

The dynamics is given by the Hamilton's equation 
\ba\label{pdot}
\dot p &=& \{p,\H\} = -\{c,p\} \f{\partial \H}{\partial c} =
2\ \g^{-1}  {\sqrt p}  \, (c-\pmk)~.
\ea
Since the Hamiltonian is a constraint, namely
$\H\approx0$,
we  find that
\be
(c-\pmk)^2= \frac{8\pi G}{3}\,\g^2 \,p\,\rho\, 
 - \, k\g^2 ~.
 \ee
We can now recover the usual Friedmann equation as
 
\be\label{Friedmann}
H^2 = \left(\frac{\dot a}{a}\right)^2= \left(\frac{\dot p}{2p}\right)^2 = \left(  \f{c-\pmk}{\g\sqrt p} \right)^2 =  \fs{8\pi G}3 \,  \rho - \frac{\pmk}{p}
\ee
\vskip11pt
In order to obtain the Raychaudhuri equation for the acceleration, we compute the equation of motion for $c$
\ba\label{cdot}
\dot c &=& \{c,\H\}  = \{c,p\} \f{\partial \H}{\partial p} \nn&=&
-\f{(c-\pmk)^2 +\pmk\g^2}{2 \g \sqrt p}\,  +
\frac{8\pi G \g}{3} 
\f{\partial\H_m}{\partial p}  \nn
&=&  -\frac{4 \pi G \gamma}{3} \sqrt p\ (\rho +3P)
\\[1pt]\nonumber
\ea
where we have introduced the thermodynamic pressure $P$ as  the derivative of $\H_m$ with respect to the volume. 
This  gives  $
{\partial \Hm}/{\partial p} 
=  -\fs32\sqrt p\ P
$.
From \eqref{pdot} and \eqref{Friedmann} we compute
\be
 \ddot p = \frac{16\pi G}{3} \, p\,\rho -2\pmk + \frac{2}{\g} \sqrt p\ \dot c ~
\ee
and we obtain%
\footnote{Equivalently, one can obtain $\dot H$ using \,
$
2H\dot H = \fs{8\pi G}3\dot\rho + 2H\f\pmk{p}
$
and 
$
\dot\rho = -3H(\rho+P)
$. 
The conservation equation will hold also in the case with quantum gravity modifications,
since it results from the Hamilton's equation for the matter part, without involving the gravitational part. }
\be
\dot H = \f{\ddot p}{2p} = \fs{4\pi G}3\,  (\rho-3P) - \f \pmk p
\ee
that yields  the Raychaudhuri equation
\be\label{ray}
\f{\ddot a}{a} =
 \dot H + H^2 = 
 -\fs43\pi G(\rho +3P)   ~.
\ee
Note that a divergence in $\rho$ and/or $P$ can lead to a divergence in the Ricci scalar 
\be\label{Ricci}
R = 6 \left(\dot H + 2H^2  +\f\pmk {p} \right)~.
\ee
Analogously, a divergence in $\dot P$, the derivative of the pressure, can lead to a divergence in
the derivative of the curvature
\be\dot R = 6 \left(\ddot H +
4 H \dot H
-2H\f\pmk{p}
\right)~.
\ee
Divergences can similarly be computed  for the higher derivatives.

\vskip10pt
\subsection{Phenomenological model}

Different types of singularities are categorized  depending upon whether the divergences appear in the scale factor $a$, in the energy density $\rho$, in the pressure $P$ or in its derivative $\dot P$.
For our investigation we choose a specific expression for the pressure
\be\label{not_pressure}
P = - \rho  - \f{A B \rho^{2 \alpha - 1}}{A \rho^{\alpha - 1} + B} 
\ee
that allows to obtain various singularities by varying the parameters of the model $A$, $B$ and $\alpha$ \cite{not}.
The derivative of the pressure is given by the expression%
$$\hspace{-1mm}
\dot P =  3H(\rho + P)
 \bigg[1  +  \f{(2 \alpha \!-\! 1) \, A B \, \rho^{2 \alpha - 2}}{A \rho^{\alpha - 1} + B}  + \f{(1 \!-\! \alpha) A^2 B \rho^{3 \alpha - 3}}{(A \rho^{\alpha - 1} + B)^2} \bigg] .$$
\vspace{-8mm}
\be\ee\\[-7mm]
The equations of motion of the model can be integrated and the scale factor can be expressed as a function of the energy density $\rho$ as
\be\label{scale_factor}
a = a_o \, \exp\left( \f{(2 A + B \rho^{2}) \rho^{1 - \alpha}}{6\,A B (1 - \alpha)}\right)~.
\ee
Here $a_o $ is an integration constant, and for the singularities at finite scale factor, it corresponds to the value where the singularity appears.
Inverting this expression we obtain
\be
  \rho = \left(-\f{A}{B} \pm 
 \sqrt{
 \f{A^2}{B^2} - 6 (\alpha - 1) A \ln \left(\f{a}{a_o}\right)
 }~
 \right)^{1/(1 - \alpha)} \!\! \!\!  \!\!  .
\ee
The following table summarize the relation between
the parameters of the model and the quantities that diverge,
while the other quantities remain finite.
 \begin{table}[htdp]\hspace{-5pt}
\begin{tabular*}{0.49\textwidth}{@{\extracolsep{\fill}} |l|c|cc|}
\hline
singularity & divergences & \multicolumn{2}{c|}{parameters}
\\ \hline 
\hline 
~Type I   
& $a\to   \infty$,~$\rho\to   \infty$,~$P\to   \infty$~ &	$3/4<\alpha<1$      & $\forall A,B$
\\[4pt] \hline
~Type II     
&	
$P\to   \infty$ &	 \multicolumn{2}{c|}{$\alpha<0$	\,~or~\,    $A/B>1$}
\\[4pt]  \hline
~Type III \,   
&	$\rho\to   \infty$ ,~~$P\to   \infty$ 	 &	$\alpha>1$	  & $\forall A,B$
\\[4pt]  \hline
~Type IV   
&	    $\dot P\to   \infty$		&	$0<\alpha<1/2$  & $\forall A,B$
\\[4pt] \hline
\end{tabular*}
\end{table}%
\\

In the classical theory, singularities can appear in the past or in the future, depending on the choice of the parameters $A$ and $B$. If $A$ and $B$ have different sign, they always give rise to a sudden singularity. The other singularities depends on the value of $\alpha$, irrespectively on the value of $A$ and $B$. No exotic singularity appears for $1/2<\alpha<3/4$.

In Sec.\ref{numerics} we study these singularities, in particular by showing the behavior of the Hubble rate, the Ricci curvature and its derivative, since these quantities are closely related to the energy density, the pressure and its derivative.

\section{Effective dynamics in LQC}\label{quantum}

The underlying dynamics in LQC captures quantum discreteness of LQG. However, for coherent states  one can obtain an  
effective continuum spacetime description. Such an  analysis has been carried  for different types of matter and an effective Hamiltonian has been 
obtained \cite{vt,st}. The effective Hamiltonian provides effective dynamics via Hamilton's equations up to controlled higher order approximations. Extensive numerical simulations show that the effective dynamics is an excellent approximation to full quantum dynamics for states which correspond to a macroscopic universe at late times \cite{aps2,apsv,aps3}.
The non-local quantum gravitational effects originating from the underlying quantum geometry primarily modify the gravitational part of the 
Hamiltonian. The matter part of the Hamiltonian constraint remains unaffected. (We discuss the underlying approximation later in a remark).

The effective Hamiltonian can be written as \cite{aps2,apsv,kv,szulc,vt,st}:
\be\label{heff0} \heff := \f{A(v)}{16
\pi G} \bigg[\sfnsqc - k\chi \bigg] + \, \, \hcalm
\ee
where 
\be
\bar \mu^2 p = 4 \sqrt{3} \pi \gamma \lp^2 =: \Delta
\ee
and $\chi$ has different expressions for $k=1$ and $k=-1$ geometries.
For $k=1$ it is given by \cite{apsv}
\be\label{emme} \chi:= 
\sfnsq - (1 + \g^2) \bar \mu^2\ee
and for $k=-1$ it becomes \cite{kv,szulc}
\be
\chi:= -\g^2 \bar \mu^2 ~.\ee

In above equations, $\Delta$ denotes the minimum eigenvalue of the area operator in LQG and $v$ denotes the eigenvalues of the volume operator $\hat V  = \hat p^{3/2}$ in LQC.\footnote{Note that the value of $K$ is different from various previous works. This is due to change in the expression for the area gap in LQC \cite{aw:bianchi1}.
In comparison to Ref. 
\cite{aps2}, we have $\bar \mu^2 p^2 = 4 \sqrt{3} \pi \g \lp^2$.} Without any loss of generalization we restrict ourselves to the positive eigenvalues of the volume oeprator.
\be\label{voleigen}
V = p^{3/2} = a^3 = \left(\frac{8 \pi \g}{6}\right)^{3/2} \frac{v}{K}
\ee
with $K = 2/3\sqrt{3\sqrt{3}}$.

For $v > 1$, the expression for $A(v)$ yields \cite{aps2}
\be
A(v) = -\frac{27 K \lp}{2 \g^{3/2}} \sqrt{\frac{8 \pi}{6}} |v| = - \frac{6 p^{1/2}}{\bar \mu^2 \g^2} 
\ee
where in the last step we have used Eq.(\ref{voleigen}). Thus, the effective Hamiltonian becomes
\be\label{heff}
\heff = -\frac{3}{8 \pi G \g^2}\frac{\sqrt{p}}{\bar \mu^2} \left(\sin^2(\bar \mu(c - k)) - k \chi\right) + \hcalm ~.
\ee
The vanishing of the Hamiltonian constraint, $\heff \approx 0$, leads to
\be
\sin^2(\mb(c - k)) = \frac{8 \pi G}{3} \frac{\g^2 \mb^2}{\sqrt{p}} \hcalm + k \chi = \frac{\rho}{\rcr} + k \chi
\ee
where we have defined the critical energy density $\rcr$ as
\be
\rcr = \frac{3}{8 \pi G \g^2 \Delta} ~.
\ee

The modified Friedmann equation can be obtained from  the effective Hamiltonian (\ref{heff}), by computing the Hamilton's equation for $p$:
\be
\dot p = \{ p, \heff\} = \f{2}{\g\mb} \sqrt p\  \sk\ck ~.
\ee
The Hubble rate $H^2 = \dot a^2/a^2$, then becomes

\ba\label{Fq}
H^2 &=& \left(\f{\dot p}{2p}\right)^2 = \f{1}{\g^2\Delta} \smk \big(1-\smk\big)\nn
&=&
\left( \f83\pi G \rho + \frac{\pmk \chi}{\g^2\Delta} \right)\left( 1- \f{\rho}{\rcr} -\pmk \chi \right) ~.
\ea
This equation can be rewritten as
\be
H^2 = \frac{8 \pi G}{3} \, (\rho - \rho_1)\left(\frac{1}{\rcr} (\rho_2 - \rho)\right)
\ee
where $\rho_1$ and $\rho_2$ are defined as 
\be
\rho_1 := \frac{- 3 k \chi}{8 \pi G \gamma^2 \Delta} = - k \, \chi \, \rcr
\ee
and
\be
\rho_2 := \rcr(1 - k \chi) ~.
\ee
In the classical limit, $\Delta \rightarrow 0$, we obtain
\be
 \chi \rightarrow -\gamma^2 \bar \mu^2,~~~ \rho_1 \rightarrow  \frac{3 k}{8 \pi G p} ~~ \mathrm{and} ~~ \frac{1}{\rcr}(\rho_2 - \rho) \rightarrow 1~.
\ee
Thus we recover back the classical Friedmann equation in the limit $\Delta \rightarrow 0$.

Similarly, using the Hamilton's equation for $c$, we can obtain the modified Raychaudhuri equation:
\ba\label{rai}
\frac{\ddot a}{a} &=& \nonumber -\frac{4 \pi G}{3} (\rho + 3 P) + \frac{16 \pi G}{3} \left(\frac{\rho}{\rcr} + k \chi\right)\left(\rho + \frac{3}{2} P \right) \\
&& + \frac{k \chi}{\gamma^2 \Delta} \left(\frac{\rho}{\rcr} + k \chi\right) - \frac{2 \zeta k}{\gamma^2 \Delta} \left(\frac{\rho}{\rcr} + k \chi - \frac{1}{2} \right)
\ea
where
\be
\zeta := \sin^2(\bar \mu) - \bar \mu \sin(\bar \mu)\cos(\bar \mu) ~.
\ee
Using eq.\eqref{rai} and eq.\eqref{Fq} we obtain the equation for the rate of change of the Hubble rate
\be\label{dotH}
\dot H = \left(-4 \pi G(\rho + P) + \frac{k (\zeta - \chi)}{\gamma^2 \Delta}\right)\left(1 - 2 \left(\frac{\rho}{\rcr} + k \chi\right)\right) ~.
\ee
It is then straightforward to verify that the conservation law 
\be
\dot \rho + 3 H (\rho + P) = 0
\ee
remains unchanged in the effective dynamics of LQC. 

We can now write down the expression for the Ricci scalar which captures the complete behavior of the spacetime curvature for the homogeneous and isotropic model:
\ba\label{ricci}
R &=& \nonumber 6\left(H^2 + \frac{\ddot a}{a} + \frac{k}{a^2} \right)\\
&=& \nonumber 6\Bigg[\frac{4 \pi G}{3} (\rho - 3 P) + \frac{8 \pi G}{3} \left(\frac{\rho}{\rcr} + k \chi\right) (\rho + 3 P)  \\
&& ~~ + \frac{k \chi}{\gamma^2 \Delta} - \frac{2 \zeta k}{\gamma^2 \Delta} \left(\frac{\rho}{\rcr} + k \chi - \frac{1}{2}\right) + \frac{k}{a^2}\Bigg] ~.
\ea

Further, as for the case of
modified Friedmann equation, the modified Raychaudhri \eqref{rai}, $\dot H$ \eqref{dotH} equations and Ricci scalar go to the classical GR versions in the limit $\Delta \rightarrow 0$. 

 It is important to point out a notable difference between the Ricci scalar in the classical theory and LQC. In the classical theory, the expression for $R$ is 
{\it independent} of the curvature index $k$ 
\be
R = 8 \pi G(\rho - 3 P) ~.
\ee
Whereas in LQC, Ricci scalar is {\it dependent} on the curvature index. The expression for the Ricci scalar is {\it different} for flat, open and closed 
models in LQC. We will see in the next section that it leads to interesting distinctions for singularity resolution for different spatial geometries in LQC.\\

We conclude this section with the following remark:\\

\noindent
{\bf Remark:} In the derivation of effective equations we have worked under the approximation that contributions from the inverse scale factor effects 
are negligible.  The approximation is well motivated due to two reasons. Firstly, from  insights gained from various numerical simulations which show that in LQC, modifications originating from the non-local field strength of Ashtekar-Barbero connection, which results in the trigonometric function in eq.(\ref{heff0}), significantly overwhelm the modifications originating from inverse scale factor \cite{aps2,apsv,kv}. Secondly, for $v > 1$ inverse scale factor effects are in any case negligible (in the fundamental representation of the theory\footnote{The situation changes when one is considering higher representations where inverse scale factor effects can lead to novel 
phenomenological effects (see for eg. Refs. \cite{cmb,effectivestate,thermal}). See, however Ref. \cite{perez} for subtleties in dealing with higher representations.}).  This corresponds to $a > 1.5 \lp$ from  \eqref{voleigen}. In our analysis we will only consider scale factors large than this value. Thus in our analysis it will be safe to make this approximation. We 
expect the results to hold true in general as inverse scale factor effects tend to weaken the strength of gravity (or make it effectively more ``repulsive''), thus aiding 
singularity resolution.


\section{Phenomenological model: numerical results} \label{numerics}

The phenomenological model introduced in Sec. IIB
allows us to study different kinds of singularities by choosing appropriate values for  various parameters in Eq. (\ref{not_pressure}).
If $A$ and $B$ have a 
 different sign, then there is always a sudden singularity
independently of  the value of $\alpha$. 
%
%
If $A$ and $B$ have the same sign, then we obtain the following classification:
\\[-14pt]
\noindent
\begin{description}
\item Type I singularity (Big Rip): $A > 0$ and $3/4<\alpha<1$.   \\[-14pt]
\item Type II singularity (Sudden): $\alpha < 0$   \\[-14pt]
\item Type III singularity (Big Freeze): $\alpha > 1$   \\[-14pt]
\item Type IV singularity (Big Brake): $0<\alpha<1/2$    \\[-14pt]
\end{description}

If $\alpha = 0$ or $3/4 < \alpha < 1$, there are no type I-IV singularities. Further, apart from the type I singularity which occurs with an associated divergence in the scale factor, type II-IV singularities can occur both in the past or the future of an expanding branch in the classical FRW model.\\

Using the effective loop quantum dynamics as elaborated in the previous section, we carried out extensive numerical simulations for various parameters  with the phenomenological equation of state (\ref{not_pressure}). Below we discuss various exotic singularities and show the representative  results for different cases. (We use Planck units to show results in various plots).


\subsubsection{Type I singularity: The Big Rip}
A type I singularity is also called ``Big Rip'' because 
there exists a finite time in which the scale factor, energy density and pressure diverge,
tearing apart  the universe.
The dominant energy condition is broken and the equation of state converges to $w=-1$ 
when approaching the singularity.

\begin{figure}[tbh!]
\includegraphics[scale=0.7]{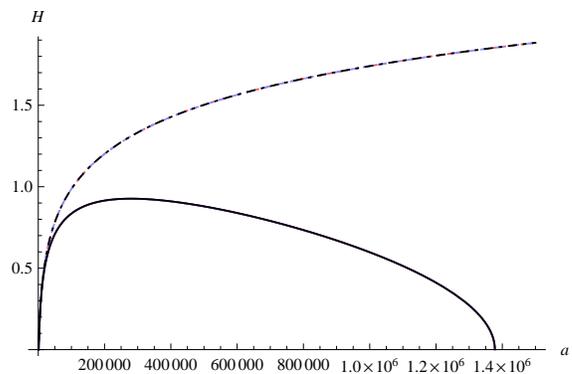}
\caption{{\bf Type I future singularity:} \ Hubble rates for  \mbox{$k=0,\pm 1$} models are shown. Dashed curves corresponds to the classical theory and the solid ones to LQC.  (Curves for different 
values of curvature index coincide with each other for the resolution  in the plot). 
In the classical theory, the Hubble rate diverges in a finite time with $a \rightarrow \infty$. We see that the effective LQC  curves behave similarly to the classical  ones
for small values of the scale factor $a$ when the energy density is very small compared to the Planck scale. As the energy density increases, departures
from classical trajectories became significant and we see a
quantum recollapse.
~In this figure the parameters are $a_o=1000$, $A = 0.1, B = 1$ and $\alpha = 0.8$. 
 }
\label{qnf1}
\end{figure}

\begin{figure}[tbh!]
  \centering
\includegraphics[scale=0.7]{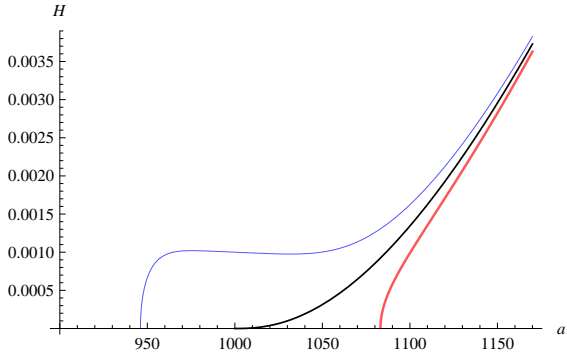}\hspace{2em}
\caption{Zoom of the Hubble rate around $a_o=1\,000$. 
  The curves are respectively  thin (blue) for $k=-1$ on the left, black for $k=0$ and thick (red) for  $k=+1$ on the right. The parameters are the same as in \fref{qnf1}. This zoom reveals the differences for evolution in Hubble rates in the region around $a_o$, that corresponds in this case to the ``initial point'' of cosmic evolution.
  }\label{qz10}
\end{figure}

In LQC, the energy density is bounded by a maximal value, therefore all type I singularities are resolved by the quantum theory.  The energy density grows as in the classical theory as the singularity is approached, but when it reaches close to $\rcr$, quantum effects lead to significant modifications to the dynamical trajectory.  The acceleration $\ddot a$ becomes negative and the Hubble rate goes to zero (\fref{qnf1}). Instead of ripping apart in finite time, the loop quantum universe recollapses and the evolution continues. 
The presence of a maximal density affects also the curvature invariants. In particular the
Ricci scalar and its derivatives remain bounded during the whole evolution.

\begin{figure}[tbh!]
\includegraphics[scale=0.7]{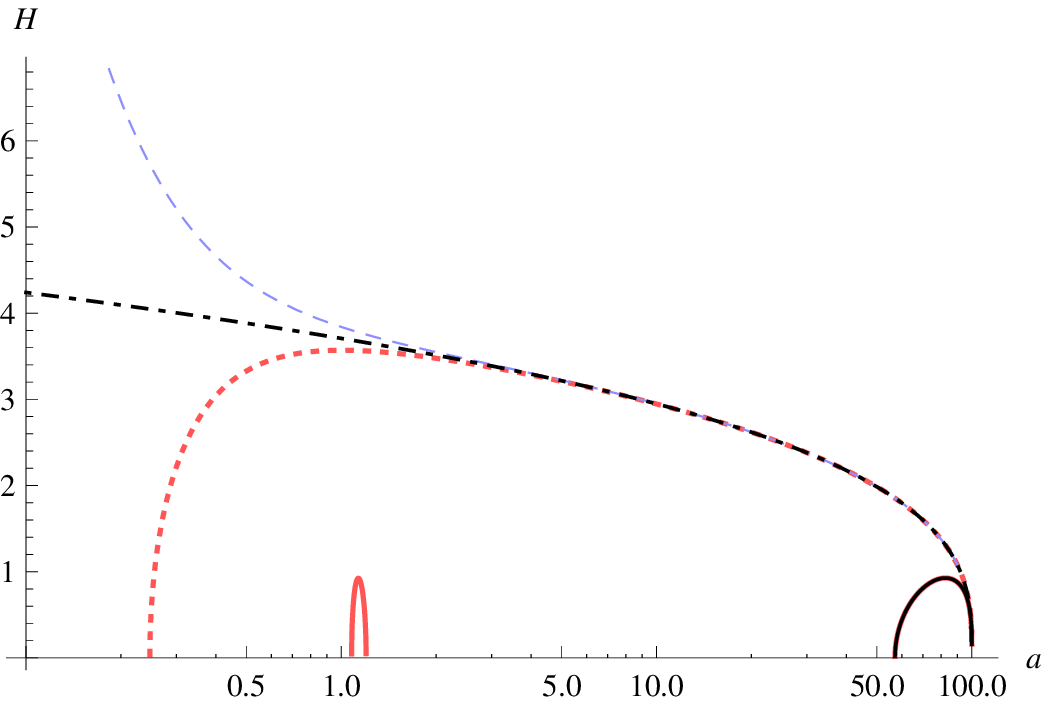}
\includegraphics[scale=0.7]{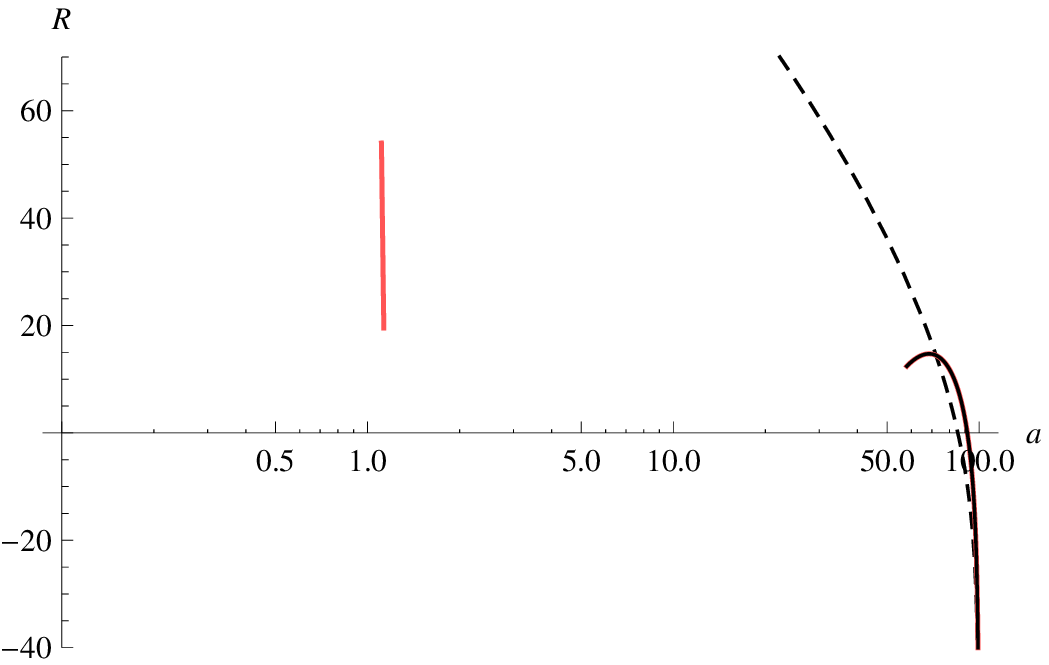}
\caption{{\bf Type II future singularity:} \ Hubble rate and  Ricci scalar for $k=0$ in black, $k=+1$ in thick (red) and $k=-1$ in thin (blue) curves are shown. Classical curves are dashed and LQC curves are solid. The Hubble rate goes to zero both in the classical theory and LQC for all values of the curvature index at $a=a_o=100$. The Ricci scalar diverges at $a_o$ in the classical theory and LQC. (Since Ricci scalar is independent of the value of curvature index in the classical model, there is only curve for the classical theory). For $k=+1$ 
there exist two disjoint solutions of the Friedmann equation, and the further branch appearing there
is bounded both in the Hubble rate and in the spacetime curvature.
~The parameters are $A = -0.1, B = 10$ and $\alpha = 1/4$.  } 
\label{qnf2}
\end{figure}

This behavior which was first noted in $k=0$ model \cite{generic} (see also Ref. \cite{phantom,phantom1} for earlier results) remains essentially the same in the $k=1$ and $k=-1$, as is shown by \fref{qnf1}. This is not surprising because type I singularities occur in future in the expanding branch where the scale factor is very large and the effects due to intrinsic curvature are expected to be negligible. \mbox{(We will} see below that effects due to intrinsic curvature can indeed produce surprising results for other exotic singularities).

The only notable difference between the $k=0$ and $k=\pm1$ cases is that the exact value of the density at the bounce changes, but the correction is small (of the order of $\lp^2/a^2_{min}$) and has no effect on the qualitative behavior of the trajectory. 
 The three cases differ only around the value $a_o$ of the scale factor as shown in \mbox{\fref{qz10}.} As one would expect for this equation of state, at large scale factors 
 the evolution in $k=\pm1$ model mimics that in the flat model and the type I singularity is resolved.

\begin{figure}[tbh!]
\includegraphics[scale=0.7]{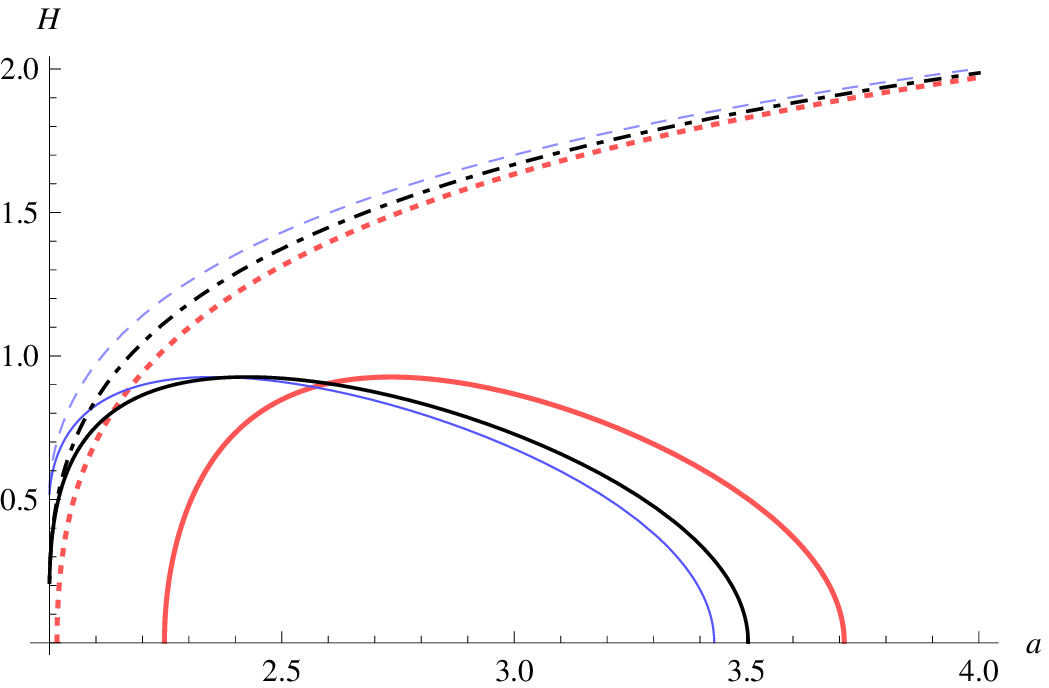}
\vskip8pt
\includegraphics[scale=0.7]{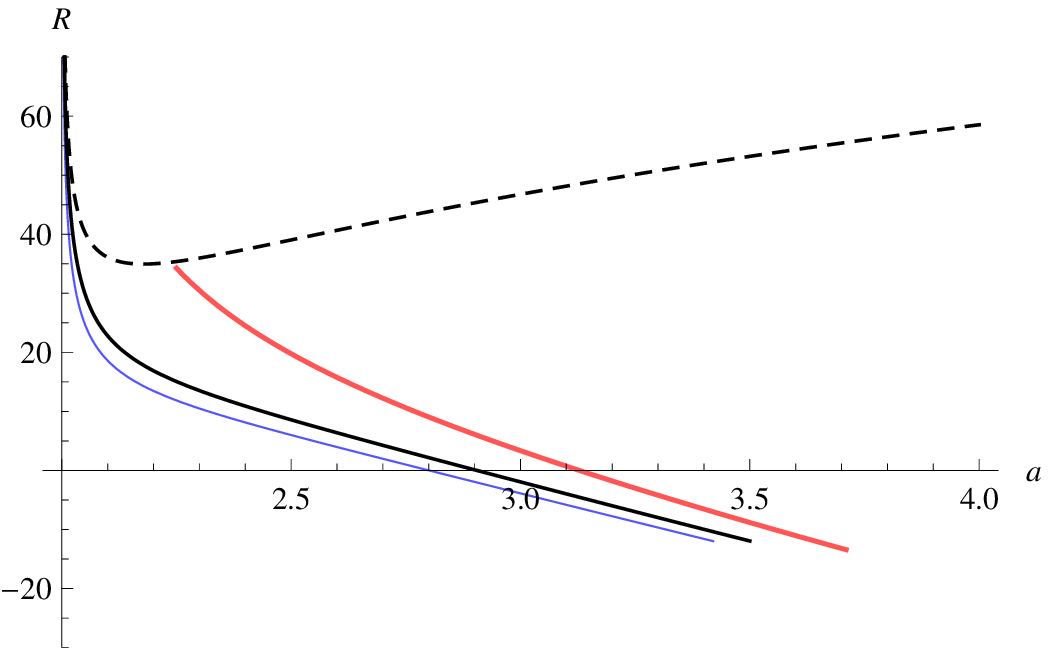}
\caption{{\bf Type II past singularity:} \ Hubble rate and  Ricci scalar for $k=0$ and $k=\pm 1$ are shown. Classical curves are dashed and LQC curves are solid.
$k=0$ is in black, $k=+1$ is thick (red) curve and $k=-1$ in thin (blue) curve.  The singularity is at $a_o=2$, where the Ricci scalar diverges. Note that for $k=+1$  the universe begins at $a>a_o$, and the Ricci scalar is always finite.
~In this figure the parameters are $A = 0.1, B = -10$ and $\alpha = 1/4$.  \\[-4pt]} 
\label{type2past}
\end{figure}

\subsubsection{Type II singularity: The sudden singularity}

In the type II singularity, the energy density does not diverge. The singularity is caused by a divergence in the pressure which results in a divergence in spacetime curvature at a finite value of the scale factor. To understand this singularity in more detail, let us  
consider Eq. \eqref{not_pressure}:  the pressure diverges when
the energy density is
$
\rho =\rho_s \equiv \left(-A/B\right)^{-\frac{1}{\alpha -1}}
$. 
Inserting this expression into \eqref{scale_factor} we find that  the singularity appears for $a\,\rightarrow\,a_o$. It is important to note that in LQC, quantum geometric effects do not regulate any divergence in pressure and the spacetime curvature can diverge \cite{generic}. It turns out that such a divergence for the type II singularity does not signal the end of spacetime. 
Geodesics can be extended beyond this
singularity. Sudden singularity also turns out to be a weak singularity because the tidal forces are not strong enough to cause a complete destruction of arbitrary detectors \cite{generic}. Thus weak singularities signal neither a break down of the underlying theory nor the end of spacetime.

In \fref{qnf2}, we show the evolution of the Hubble rate for different values of 
spatial curvature index for type II singularity occurring in the future of an expanding branch. We see that  both in the classical theory (dashed curves) and LQC (solid curves), the Hubble rate vanishes in a finite time for all values of the spatial curvature at scale factor $a = a_o$. In contrast, the  plot of the Ricci scalar shows that it diverges
at this value of the scale factor for all values of the spatial curvature in the classical theory as well as in LQC.

As is clear from \fref{qnf2}, the results about future sudden singularity are on expected lines of the $k=0$ model when the scale factor is very large, where the effects due to intrinsic curvature become negligible. However, a surprising result emerges for the $k=1$ model. Our numerical analysis shows that, 
for certain values of the parameters, 
there exists an additional baby evolutionary branch separated from the main evolutionary branch. For $k=+1$,  $H^2$ is positive in two disjoint intervals. Consequently
there exist two disjoint branches corresponding to real solutions of the Friedmann equation. The extra branch occurs at small scale factors and signifies the non-trivial new physics which emerges from the quantization of intrinsic curvature in LQC. The new branch, that we can see in  the \fref{qnf2}, 
is bounded both in the Hubble rate and in the spacetime curvature.

\begin{figure}[t]
\includegraphics[scale=0.7]{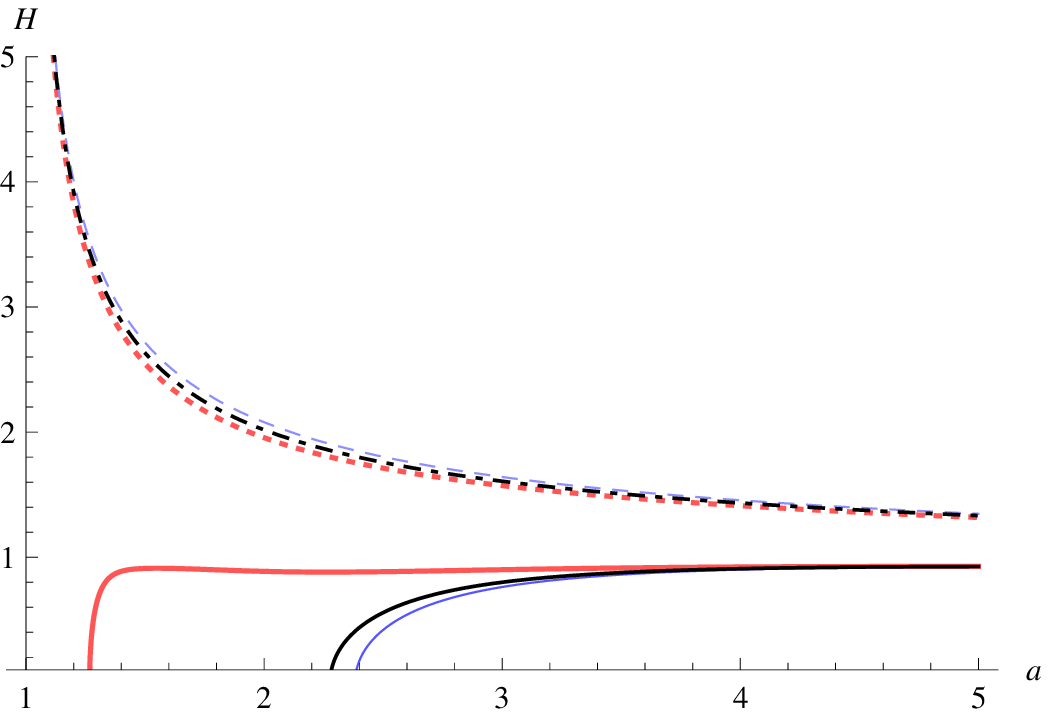}
\includegraphics[scale=0.7]{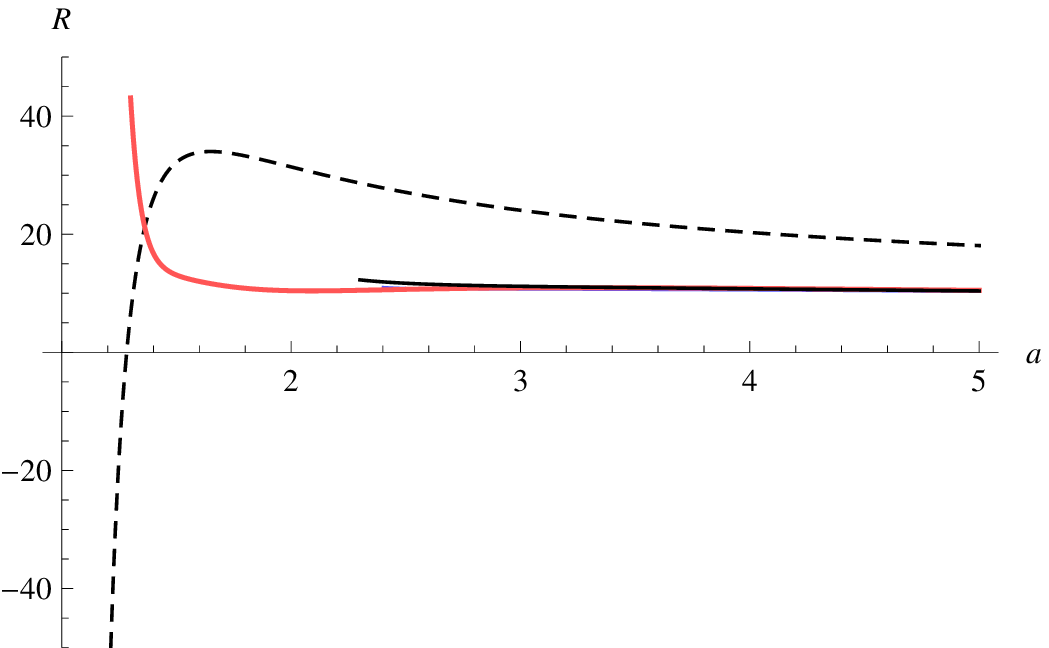}

\caption{{\bf Type III past singularity:} \ Classical (dashed) and effective LQC (solid) Hubble rate and Ricci scalar are shown.  The curves are respectively thick (red) for  $k=+1$ starting on the left, black for $k=0$ in the middle and thin (blue) for $k=-1$ stating a bit more on the right.  In the classical case there is a divergence of both $H$ and $R$ at $a_o=1$ while the curves for LQC remain bounded (in the figure the Ricci scalar for $k=-1$ is covered by the black line of the $k=0$ case).
~Here the parameters used are $A = -100, B = -1$ and $\alpha = 2$. } 
\label{type3past}
\end{figure}

Another interesting feature which can be seen from \fref{qnf2} is that for $k=1$ model, the classical curve (shown with a dotted curve) depicts that 
there is no initial singularity. The Hubble rate vanishes both in the past and the future of the classical evolution. The classical universe faces a sudden singularity in future, but since it is a weak singularity, the classical $k=1$ universe is geodesically complete for the considered equation of state. This does not hold true for the $k=0$ and $k=-1$ case, and these classical spacetimes are past incomplete.

\fref{type2past} shows the plot of Hubble rate and Ricci scalar for different values of curvature index when the sudden singularity occurs in past. In such a case the effects due to intrinsic curvature lead to novel features in the physics of singularity resolution. We find that classically the Hubble rate vanishes for all values of $k$ at $a=a_o$ and the Ricci scalar diverges. However in LQC, though the Hubble rate vanishes for values of $k$, the 
divergence in Ricci scalar does not occur for a closed universe. It turns out that as scale factor approaches $a_o$, the Ricci scalar increases to large values but it does not diverge for the allowed values of the scale factor. (The divergence occurs in the forbidden region of effective LQC dynamics where the Hubble rate is imaginary).
This result implies that in the closed model for the parameters considered here, quantization of intrinsic curvature leads to a resolution of the past sudden singularity. This is a direct consequence of the way intrinsic curvature terms enter in to the expression of the Ricci scalar in LQC. 
\mbox{We emphasize} that type II singularities whether in future or past, even though they are harmless, are not resolved in LQC for $k=0$ and $k=-1$ models. In contrast we find that these singularities are resolved, for a certain choice of parameters, when occurring in the past for $k=1$ model.


\subsubsection{Type III singularity: The Big Freeze}

\begin{figure}[b]
\includegraphics[scale=0.7]{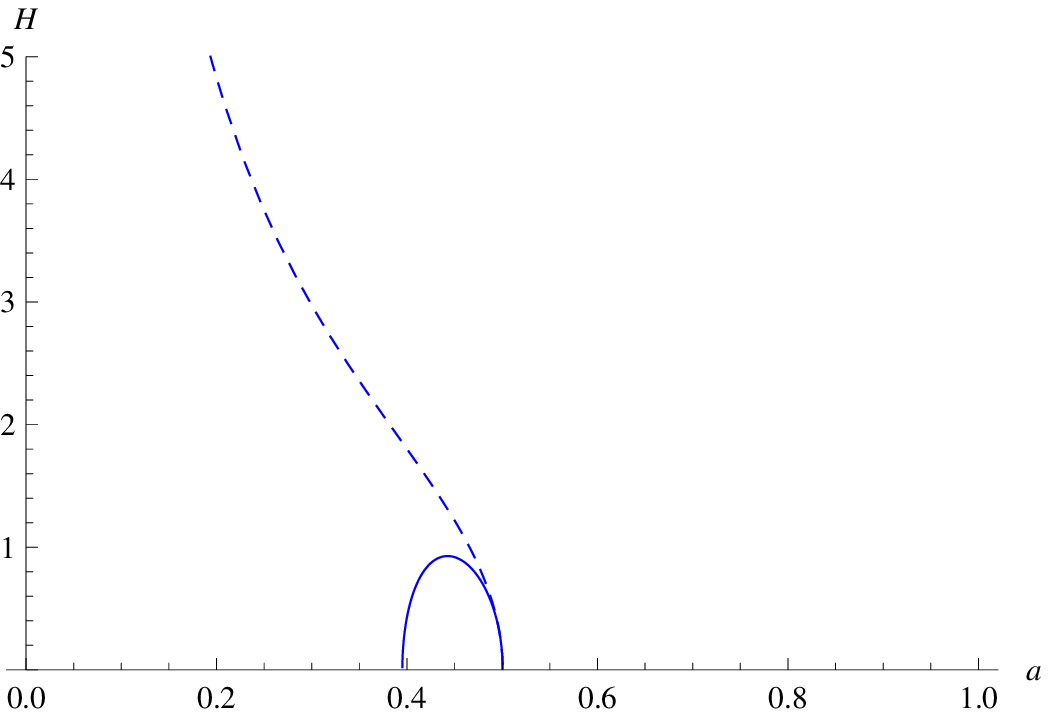}
\includegraphics[scale=0.7]{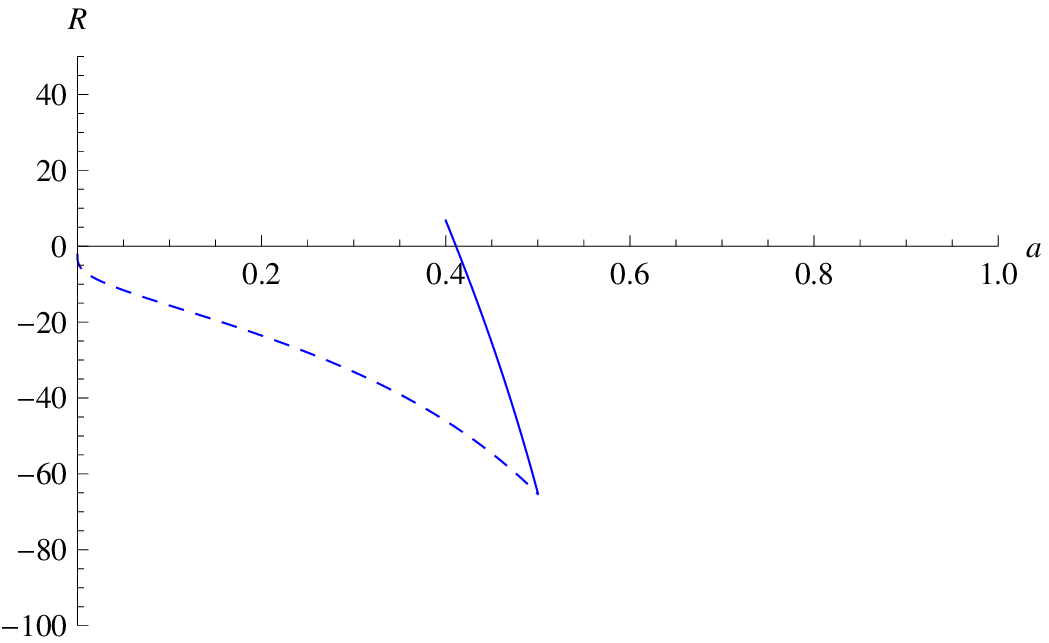}

\caption{This figure shows the additional branch occurring at small scale factors in $k=-1$ model for equation of state leading to a type III past singularity in Fig. \ref{type3past}. The additional branch occurs for certain values of parameters both in the classical theory (dashed curve) and LQC (solid curve) when scale factor is below the value $a_o$, where the big freeze occurs classically. Note that, also in this case, only LQC is immune from primordial singularity.%
~The parameters used are $A = -100, B = -1$ and $\alpha = 2$. }
\label{type3past2}
\end{figure}

\begin{figure}[t]
\includegraphics[scale=0.7]{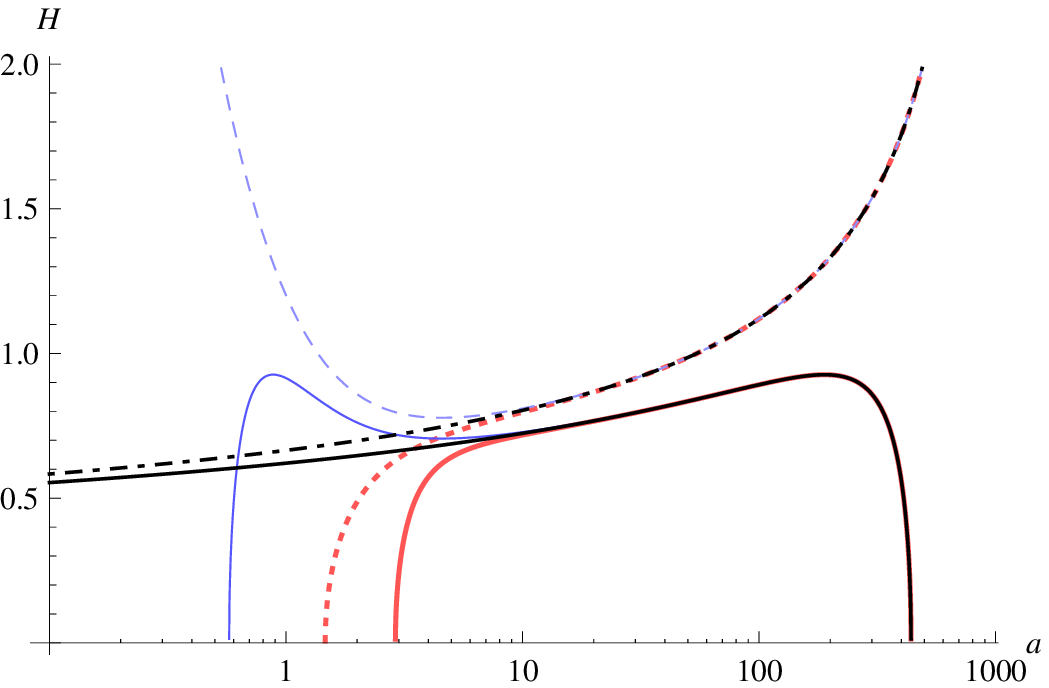}
\vskip5pt
\includegraphics[scale=0.7]{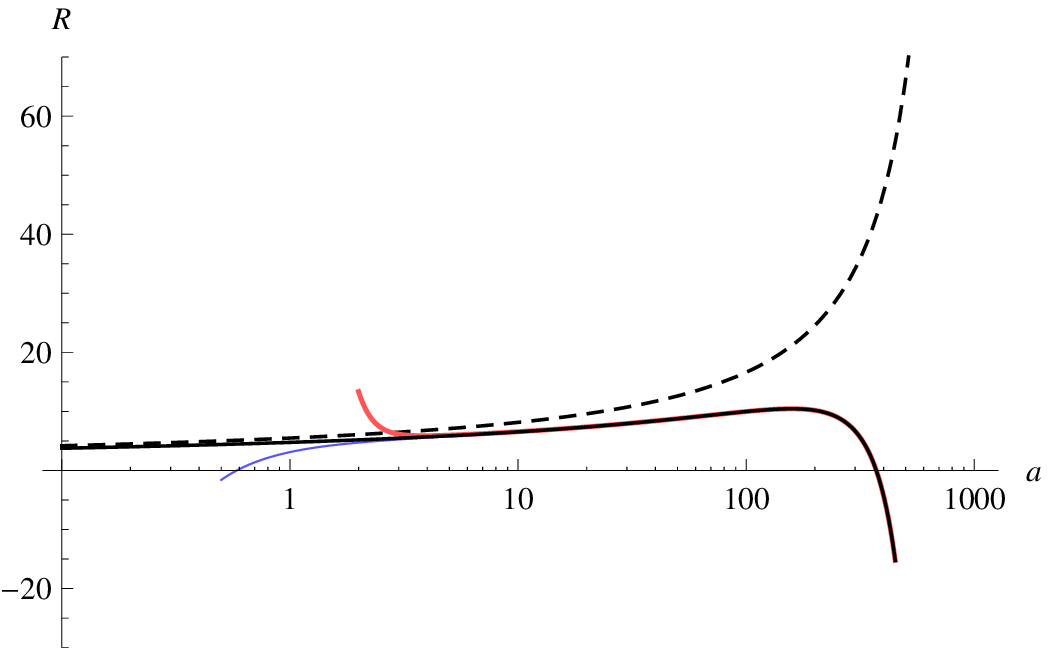}

\caption{{\bf Type III future singularity:} \ Classical (dashed) and effective LQC (solid) Hubble rate and Ricci scalar with $k=0,\pm1$ are compared. In the classical case there is a divergence of both $H$ and $R$ at $a_o$ while the curves for LQC remain bounded.
Notice that LQC cures also the initial singularity in the curve case, and in particular the thin (blue) solid curves ($k=-1$) shows a characteristic `tilt' for small $a$.%
~The parameters are $a_o=1000$, $A = 100, B = 1$ and $\alpha = 2$. }
\label{type3future}
\end{figure}

A type III singularity occurs at a finite value of scale factor where both the Hubble rate and the Ricci scalar diverge. 
In LQC, the presence of a maximal energy density $\rcr$ bounds
the Hubble rate, giving rise to a recollapse when the singularity is approached. Also $\ddot a/a$ and the Ricci scalar are bounded and finite. Thus type III singularities 
are generically resolved in LQC for all values of the curvature index.

This can be seen from the plots in \fref{type3past}, where we have shown the plots of the Hubble rate and the Ricci scalar when the big freeze singularity occurs in the past at $a=a_o$. We see that in the classical theory, Hubble rate and Ricci scalar diverge as $a \rightarrow a_o$ in the classical theory. On the other hand, in LQC the Hubble rate vanishes in the past showing that there is a bounce. Further, the Ricci scalar remains finite for all the values of the curvature index. 

We now make an interesting observation for $k=-1$ case. For a certain choice of parameters which lead to a  type III singularity in the past, we find that there exists an additional 
classical branch at the small scale factors. The additional branch faces a big bang singularity in the past of classical evolution and is free from singularity in future evolution (as is shown by the  dashed curves for the Hubble rate and the Ricci scalar in \fref{type3past2}).
However the additional branch occurs in LQC when the scale factor is less than the Planck length. Since the length scale involved is below where we expect the 
effective dynamics in  LQC to be valid, a more detailed analysis is needed, by including modifications pertaining to the inverse scale factor,  in order to understand the 
physics emerging from LQC in this special case. Never the less, if we assume the validity of the effective Hamiltonian in this regime, then LQC resolves the past singularity 
in the additional branch (as depicted by the solid curves for the Hubble rate and the Ricci scalar in \mbox{\fref{type3past2}}).

The type III singularity is also resolved for all values of the spatial curvature when it occurs in future. This is on the expected lines, as the effects due to intrinsic 
curvature become small at large scale factors. In \fref{type3future} we depict results from such an evolution. In the classical theory, there is a big freeze singularity in future where the Hubble rate and the Ricci scalar diverge. For the $k=-1$ universe, there also exists a singularity (big bang) in the past evolution.
We can see that in LQC there is no type III singularity as the Hubble rate vanishes at $a=a_o$ and Ricci scalar reaches a finite value. Further, the past big bang singularity  in the $k=-1$ case, as pointed above, is also resolved. The LQC universe is bounded for all values of the curvature index for the equation of state which leads to a type III singularity in the classical theory.

After performing various numerical simulations we reach the conclusion that type III singularities, irrespective of them occurring in the past or the future are always resolved in LQC for all values of the spatial curvature \mbox{index.}

\begin{figure}[tbh!]
\includegraphics[scale=0.7]{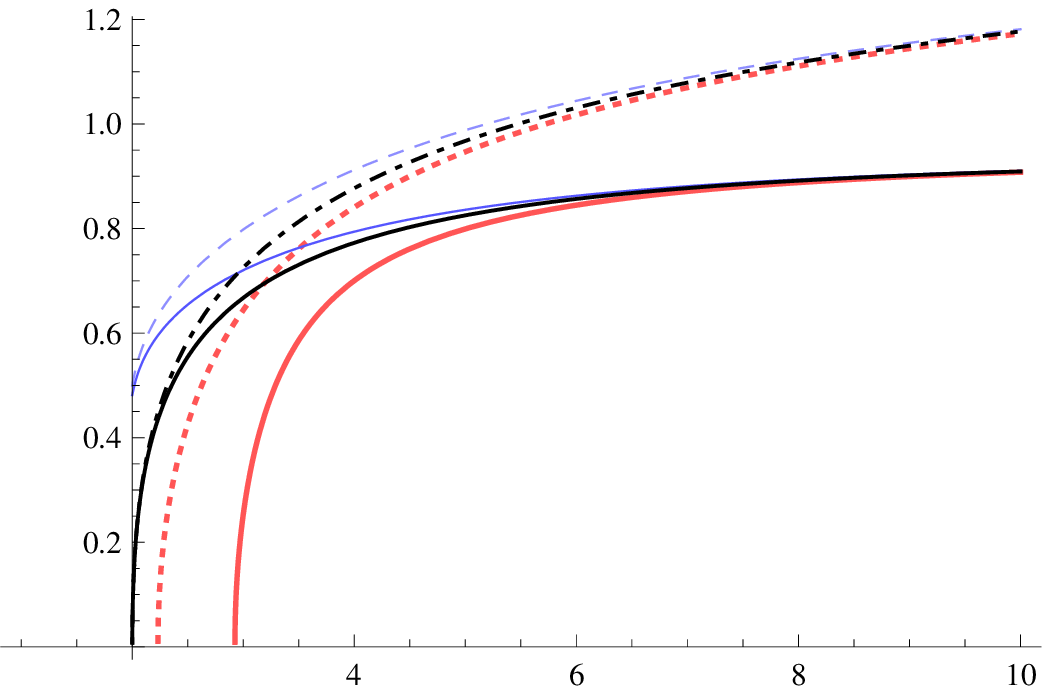}
\label{qc40}
%
\includegraphics[scale=0.7]{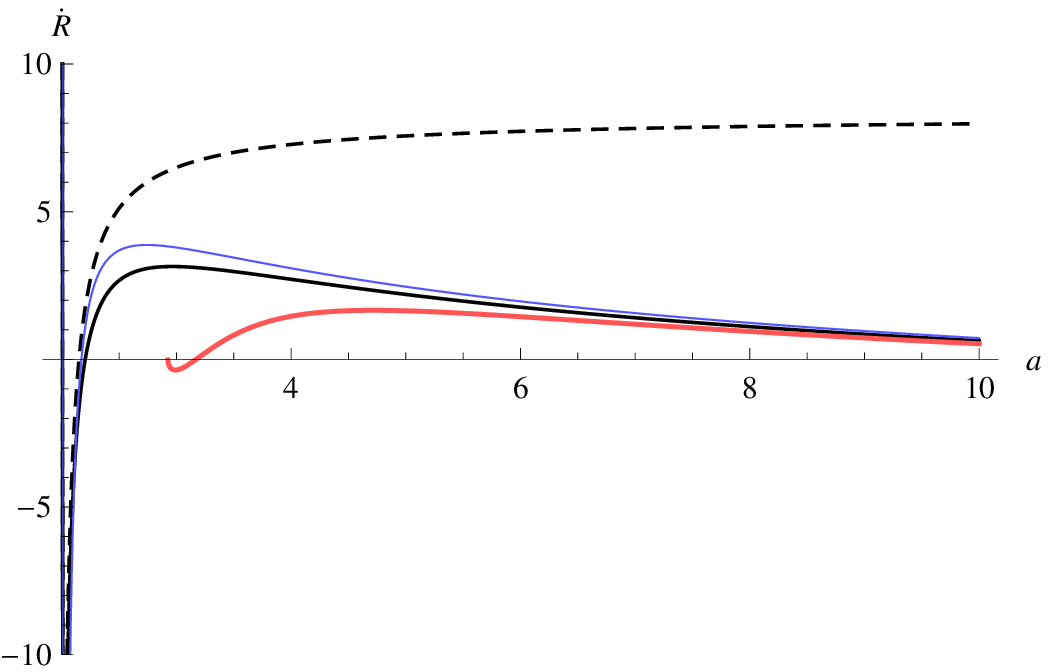}
\label{qo40}
\caption{{\bf Type IV past singularity:} \ Hubble rate and  Ricci scalar for $k=0,\pm1$. The singularity is at $a_o=2$, where the Ricci scalar diverges for $k=0$ and $k=-1$. Note that  for $k=+1$ (dotted line for the classical solutions and thick red line for LQC effective solutions) the universe begin at $a>a_o$, thus the Ricci scalar is always finite.
~The parameters are \mbox{$A = 0.01$}, $B =1$ and $\alpha = 1/4.$  } 
\label{4past}
\end{figure}

\subsubsection{Type IV singularity: The Big Brake}

Type IV singularity is a derivative-curvature-singularity where
			none of the curvature invariants diverge. (In this sense, it does not qualify as a curvature singularity). 
Though the energy density and pressure remain finite, a higher derivative of 
			the curvature diverges as $a \rightarrow a_o$. 
The value of $\alpha$ determines the order of 
			derivative which blows up \cite{not}. 
			
As in the type II case, the geodesic equations are well-behaved
since Hubble rate  is finite at $a = a_o$. 
The singularity is weak because it occurs at a finite volume and the theory is geodetically complete since the Hubble rate is bounded.
Quantum geometric effects have little influence
			on this harmless extremal event beyond which geodesics 
			can be extended even in the classical theory \cite{generic}.
 
The behavior of the curves is very similar to the type II case. 
\,
In \fref{4past} we have shown the  results from the past singularity. (We have chosen $\alpha=1/4$ so that the divergence appears in $\dot R$).
We find that the Hubble rate vanishes at $a=a_o$ both in the classical theory and LQC. However, 
$\dot R$ diverges except for the spatially closed model in LQC. This result is similar to what we obtain for type II singularity. It turns out that LQC resolves type IV weak singularity when occurring in past for the closed model for certain choices of parameters.  As in the type II case, this result highlights the surprises which quantization of intrinsic curvature may bring in comparison to earlier studies \cite{generic}. For $k=0$ and $k=-1$ model, this weak singularity is ignored by the quantum geometry for all values of the parameters.

\begin{figure}[tbh!]
\includegraphics[scale=0.7]{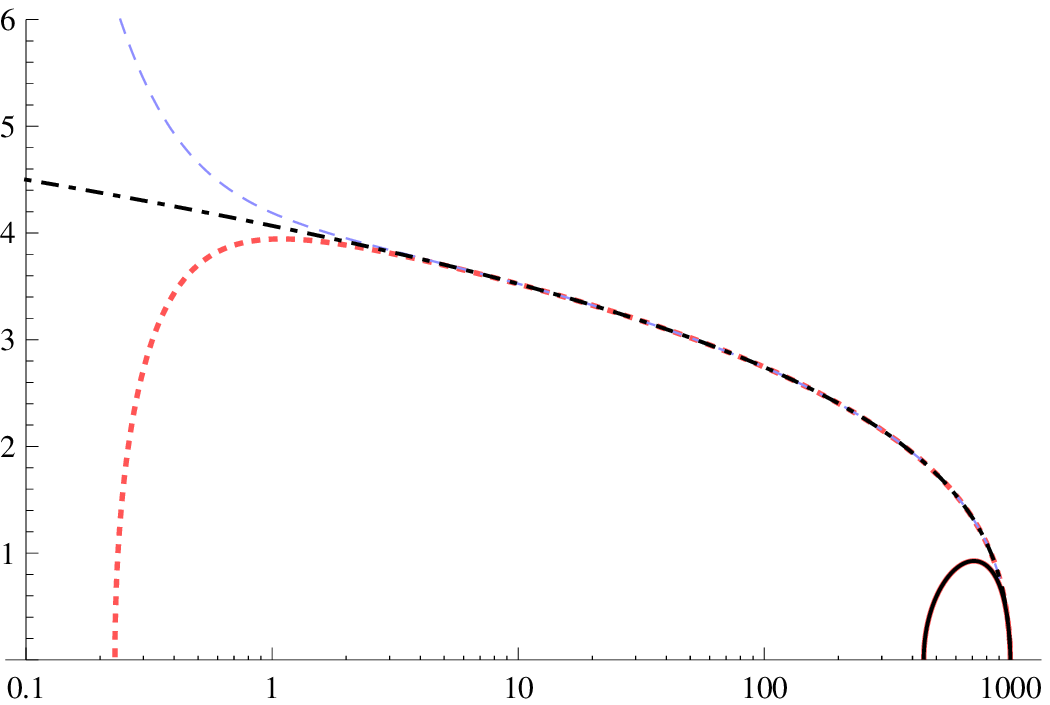}
\label{rc40}
%
\includegraphics[scale=0.7]{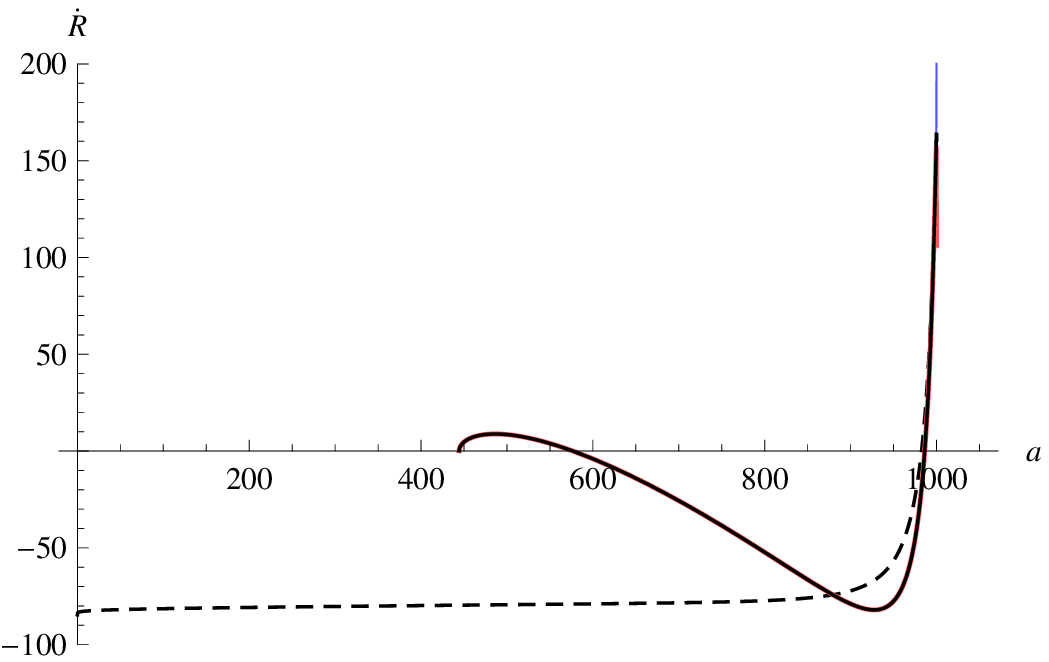}
\label{ro40}
\caption{{\bf Type IV future singularity:} \ Comparison of Hubble rates and $\dot R$ for the classical theory and LQC (in the figure LQC curves overlap). The singularity appears in the future at $a_o=1000$. The LQC universes (solid lines) do not have primordial singularity. Interestingly, for $k=1$ this happens also classically (dotted line).
In this picture there are no baby universe, but they can be obtained for $k=1$ for a different choice of the parameters.
~Here the parameters are $A = -0.1$, $B =-1$ and $\alpha = 1/4.$  }   
\label{4future}
\end{figure}

We show the results for type IV singularity occurring in future evolution in  \fref{4future}. Here we find that for all the values of spatial curvature, the 
Hubble rate vanishes at $a=a_o$ with a divergence in $\dot R$ for both the classical theory and LQC. We see that not only the LQC curves have no big bang, but this happens also classically for the closed universe (\fref{4future}, dotted line). Furthermore, for $k=+1$ it is possible to obtain a baby universe for an appropriate choice of the parameters, as  for type II singularity (see \fref{qnf2}). In this case $\dot R$ is strongly negative but bounded, thus there is no singularity in the past and in the future.

\vskip1cm

\section{Summary}\label{summary}

A fundamental question in quantum gravity is whether spacelike singularities of the classical theory are resolved. Since not all such singularities signal end of the spacetime,
it is important to understand the role of quantum gravitational effects in resolution of strong singularities (those beyond geodesics can not be extended) and 
weak singularities (those beyond which geodesics can be extended). These issues were addressed in the loop quantization of cosmological models which are spatially flat and it was found the non-perturbative loop quantum effects resolve all strong singularities and ignore weak singularities \cite{generic}.

The aim of the present analysis was to investigate these issues for spatially curved models using phenomenological model of equation of state permitting exotic 
singularities such as big rip, sudden singularities, big freeze singularity and the big brake singularity.  In order to capture the role of intrinsic curvature, we considered 
exotic singularities both in the future and the past evolution. To our knowledge, even in the classical theory exotic singularities had not been studies earlier for the spatially curved model. For the singularities occurring in the future, the contribution of the intrinsic curvature is expected to become very small and we expect results to agree with the spatially flat case. This turns out to be true. However, more interesting are cases where the exotic singularities 
occur in past. Here one would expect effects due to quantization of intrinsic curvature to play a non-trivial role. In fact we encounter some surprising results. Though strong 
singularities are always resolved in LQC, it turns out that for the closed model weak singularities occurring in the past evolution may also be resolved. Thus LQC, does not always 
ignores weak singularities. This is an intriguing result which deserves further investigation.

Another  peculiar feature of the curved models is the appearance of a small branch for 
type II and type IV singularities for certain values of the parameters. This ``baby-universe" is bounded, and is devoid of any singularities. It will be interesting to analyze 
these additional branches which we find for both spatially open and closed models in more details, in particular by taking in to account inverse scale factor effects which 
may play some role when scale factor is below the Planck length.

Our results extend earlier results on generic resolution of strong curvature singularities in spatially flat model in LQC to the spatially curved models. They also bring 
some important lessons, the primary one being that quantization of intrinsic curvature may throw some novel unexpected results and we need to gain more insights on when 
quantum gravity may ignore or resolve a weak curvature singularity. Another lesson is that as for the spatially flat model, spacetime curvature invariants may diverge 
for the spatially curved models and yet there may be no physical singularity. 

These results strengthen the case for a generic resolution of strong singularities in LQG. To achieve this goal, the next step will be to include anisotropies and then inhomogeneities. The latter will require us to go beyond the minisuperspace approximation considered here. This brings up additional challenges such as the complete 
classification of the strong and weak singularities in inhomogeneous situations. Two promising avenues where such an analysis can be undertaken would be the Gowdy models \cite{gowdy} and in the spinfoam paradigm \cite{spinfoams}. We hope that these studies will also provide insights on the deeper relation of these frameworks with LQC.

\vskip0.5cm

\section*{Acknowledgments}
\noindent
FV thanks Elena Magliaro, Frank Hellmann and Carlo Rovelli for useful discussions.




\end{document}